\documentclass[aps,pra,twocolumn,groupedaddress,superscriptaddress,tightenlines,nofootinbib]{revtex4}
\usepackage{setspace}
\usepackage[utf8]{inputenc}
\usepackage{amssymb}
\usepackage[T1]{fontenc}
\usepackage{color,graphicx}
\usepackage{lmodern}
\usepackage{amsmath}
\usepackage{amsbsy}
\usepackage{float}
\usepackage{bbm}
\usepackage{bm}
\usepackage{epsfig}
\usepackage{braket}
\usepackage{graphics}
\usepackage{caption}
\usepackage{subcaption}
\usepackage{appendix}
\usepackage{hyperref}
\usepackage{natbib}

\begin{document}

\title{Influence of Oscillating Magnetic Fields on the Electric Dipole Moment of Radical Pairs in Cryptochrome‑Based Magnetoreception.}

\author{Ali Soltanmanesh}
\email{a.soltanmanesh@gmail.com}
\affiliation{Center for Quantum Engineering and Photonics Technology, Sharif University of Technology, Tehran, Iran.}

\author{Mahboobe Sehati}
\email{mahboobesehati@gmail.com}
\affiliation{Center for Quantum Engineering and Photonics Technology, Sharif University of Technology, Tehran, Iran.}

\author{Sareh Rostami}
\email{sareh.rostami@researcher.sharif.edu}
\affiliation{Center for Quantum Engineering and Photonics Technology, Sharif University of Technology, Tehran, Iran.}

\author{Abolfazl Bahrampour}
\affiliation{Center for Quantum Engineering and Photonics Technology, Sharif University of Technology, Tehran, Iran.}

\author{Alireza Bahrampour}
\email{Bahrampour@sharif.edu}
\affiliation{Center for Quantum Engineering and Photonics Technology, Sharif University of Technology, Tehran, Iran.}
\affiliation{Department of Physics, Sharif University of Technology, Tehran, Iran.}

\begin{abstract}
Radical pairs induced by light-driven reduction of cryptochrome protein constitute a spin-dependent mechanism that is accompanied by an electric dipole moment and is found to be sensitive to external magnetic fields. In this research, to investigate for the further proof of such model, the simultaneous effect of the Earth’s static magnetic field and the time-dependent magnetic field noise on the electric dipole moment of the radical pair has been studied within the quantum mechanical framework. The effect of the external magnetic field discussed in different angles regarding the Earth magnetic field within various frequencies and magnitudes. The sensitivity of the system behavior to the external magnetic field frequencies and magnitudes, vastly differs among the changes in the magnetic field angle to the Earth's static field. Furthermore, the sensitivity studied under the effect of the environmental noise. The relative spatial orientation of the two magnetic field components plays an important role in the time evolution of the electric dipole moment. Also, deeper discussions on specific relative orientations of the external magnetic fields, such as $24^\circ{}$ degree, shows that the quantum model of radical pairs which is based on dipole moment, is in agreement with the results of the birds behavorial studies. These findings provide new insights into the sensitivity of the radical pair model to the combination of magnetic fields and may contribute to a comprehensive understanding of the phenomenon of magnetoreception and the advancement of bioinspired magnetic sensors.

\end{abstract}

\maketitle

\section{Introduction}
Various types of living organisms, from plants and microorganisms to a wide range of animals, including vertebrates and invertebrates, can sense the geomagnetic field \cite{clites2017identifying, schneider2023over}. In this regard, long-distance migratory birds have attracted the attention of many scientists as to how they can detect their current location and navigate to a specific destination. Since ancient times, talking about the magnetic field and navigation has always reminded humans of the use of a compass. Therefore, it is expected that these organisms have internal magnetic compasses. In this context, Schulten et al.,\cite{Schulten1978} proposed the radical pair mechanism to explain how the Earth's magnetic field interacts with biological structures. Various behavioral studies have been conducted to investigate this mechanism in diverse types of birds \cite{zhang2015radical,wiltschko2021magnetic,wiltschko2019magnetoreception}, insects \cite{vacha2009radio}, etc \cite{rankin2015finding, natan2017symbiotic}. Overall, the unique features of the avian compass, such as its requirement for light of a specific wavelength and its sensitivity to the inclination of the geomagnetic field lines, are in agreement with the proposed radical pair model \cite{zhang2023quantum}. In addition, models for this mechanism have been proposed based on theoretical studies. In general, magnetoreception based on the radical pair model arises from the absorption of electromagnetic waves with specific intensity and wavelength via the chromophore group in the cryptochrome protein, a flavoprotein recognized as a potential magnetoreceptor in the avian retina \cite{hore2016radical, gortemaker2022direct}. In this process, the two radicals forming a radical pair are spin-correlated, a feature that makes magnetoreception not only a biological,  but also a quantum phenomenon, which can be studied in the field of quantum biology. The spin of this radical pair can change through interconversion from singlet to triplet state and vice versa. The interconversion rate can be affected by factors such as the interaction of the electron with the nucleus and the interaction of the electron with an external magnetic field \cite{hore2016radical}. The external magnetic field can be static such as the geomagnetic field or time-dependent fields. Since \textit{in vitro} experiments have shown the effect of alternating magnetic fields on the efficiency of organic reactions \cite{Luo2024}, it is anticipated that these fields could also affect the radical pair mechanism. To investigate the effect of oscillating fields on the spin dynamics of the radical pair mechanism, various studies have been conducted in different frequency ranges, some of which are briefly mentioned below.
Timmel et al., \cite{timmel1996oscillating} used analytical and numerical methods to investigate the influence of oscillating magnetic fields on the radical pair reaction yield. Their findings showed that depending on the lifetime of the radical pair, the intensity and frequency of the applied field, the yield of products can be affected.Hiscock et al., \cite{hiscock2017disruption} used the approximate Floquet method to investigate the effect of monochromatic and broadband radiofrequency magnetic fields on the radical pair mechanism. It is worth noting that in a previous study \cite{hiscock2016floquet}, using the method and considering the radiofrequency field as a perturbation, they were able to obtain the yield of the radical pair reaction product with acceptable approximation. However, in their recent study, the results showed that the current radical pair model cannot explain the results of behavioral studies. In an experimental study conducted by Bojarinova et al., \cite{Bojarinova2020} was reported that garden warblers disorientation subjected to time- dependent magnetic fields is independent of the photochemical magnetoreception in their retina. Leberecht et al., \cite{Leberecht2022} investigated the effect of broadband radiofrequency magnetic fields on the magnetic orientation of Eurasian blackcaps. Their results showed that in the frequency range of 75–85 MHz, their orientation ability was disrupted, indicating that at a minimum one of the radicals involved in the radical pair model must have a significant number of hyperfine interactions, which is in agreement with a flavin-derived radical pair. Muheim et al., \cite{Muheim2023} investigated the effect of radiofrequency magnetic fields on the magnetic orientation of zebra finches. Their results showed that radiofrequency fields can affect the birds' magnetic sense depending on the frequency of the field (e.g., the Larmor frequency), its intensity, and the degree of familiarity the bird has acquired through training, and their findings also support a radical pair mechanism. Luo et al., \cite{Luo2024} investigated the effect of radiofrequency fields on the spin dynamics of three radical pair models. They used two methods, $\gamma$-COMPUTE algorithm and the stochastic Schrödinger equation. Their results showed that in the absence of a static magnetic field in all three cases, including the toy spin radical pair, where only one of the radicals has two nuclear spins, a simplified version of the radical pair formed in cryptochrome, and its more realistic version, the applied radiofrequency field was resonant with the spin dynamics of the radical pairs and affected the quantum yield of the products. They also suggested that experimental results based on this theoretical study could support radical pair-based magnetoreception. Based on what is briefly reported above, a clear and definitive understanding of the mechanism of magnetic signal perception under time-dependent magnetic fields is still lacking. It is evident that more extensive interdisciplinary research, both theoretical and experimental, is required in this important field, related to wildlife and human health. In the present study, we have attempted to explain how magnetic signal detection is achieved through the radical pair mechanism within the framework of quantum mechanical calculations by examining the effect of oscillating fields on the electric dipole moment as a measurable physical quantity based on the previously provided model \cite{Sehati2025}.
In this paper, 
The organization of the paper is as follows. In Section 2, the relevant theoretical model is provided. This is followed by the outcome of our numerical solution and a discussion in Section 3 and the  conclusion of the paper in Section 4.

\section{Theoretical Framework }
\begin{figure}
    \centering
    \includegraphics[width=\linewidth]{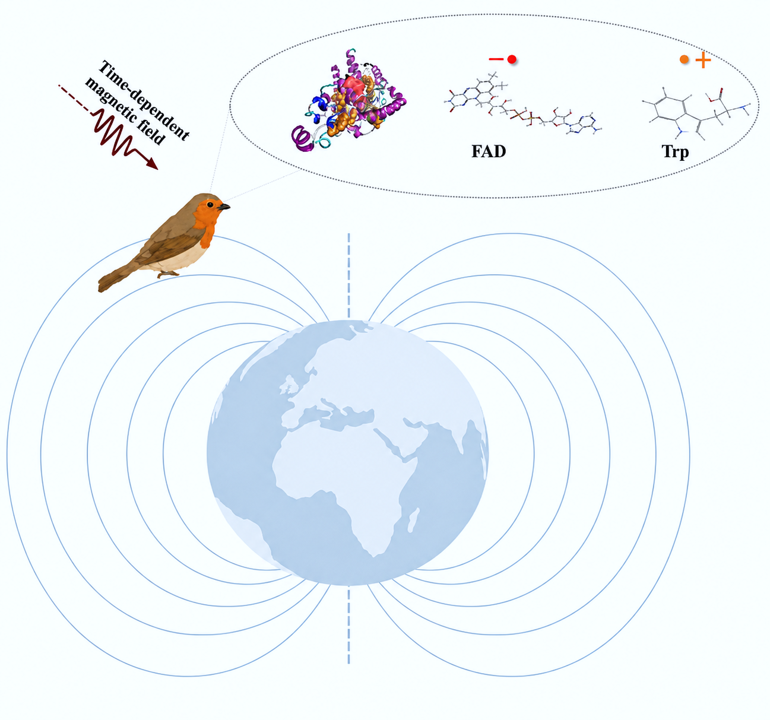}
    \caption{Schematic depiction of the photoexcitation of the cryptochrome protein, leading to the formation of a spin-correlated radical pair. The resulting charge separation gives rise to an electric dipole moment. This radical pair can be influenced by both Earth's static and external time-dependent magnetic fields.}
    \label{2nd_re_my}
\end{figure}

To investigate how external magnetic fields can modulate the mechanism of birds' internal compass, we extended our previous  work [\cite{Sehati2025}] on the dynamics of radical pairs in the protein cryptochrome (Fig\ref{2nd_re_my}). In this framework, changes in the electric dipole moment of the system have been of particular interest. The main idea is that the magnetic field alter the quantum evolution of the spin of the radical pair, and this change affects the spatial distribution of charges, ultimately leading to a change in the dipole moment. This physical quantity can serve as a biophysical pathway for converting magnetic information into understandable biological signals \cite {stoneham2012new}. The time evolution of the system is considered based on the von Neumann equation.
\begin{equation}
\frac{d}{dt} \rho(t)= -\frac{i}{\hbar} [{H}, \rho]
\label{von}
\end{equation}

where $\rho$ is the density matrix, and ${H}$ is the Hamiltonian of the system.
The Hamiltonian of the radical pair system consists of three principal terms \textit{vis-\`a-vis} hyperfine interaction, Zeeman interaction, and spin-orbit coupling, as follows:

\begin{equation}
{H} = \mathcal{I} \cdot A \cdot {S}_1 + \gamma B \cdot ({S}_1 + {S}_2) + \sum_{j=1}^{2} \zeta_j \, {L}_j \cdot {S}_j
\label{FullHamiltonian}
\end{equation}

where $ \mathcal{I}$ defines the nuclear spin operator, $\mathbf{A}$ is a diagonal anisotropic hyperfine tensor, ${S}_1$ and ${S}_2$ are the spin operators of the two unpaired electrons in the radical pair, $B$ is the external magnetic field intensity, $\gamma$ is the electron gyromagnetic ratio, $L$ and $S$ are the orbital angular momentum and spin operators respectively, $\zeta_{SOC}$ is the spin-orbit coupling constant, and indices refer to each of the radicals. Here the dirst term is introducing anisotropy to the system, the second term explains the interaction of the internal magnetic field with the electron spin and the interplay of the external magnetic field with the electron spin, respectively, regardless of the inter-radical interactions. The third term represents the spin-orbit interaction, which establishes the connection between the spin and orbital angular momentum of the electron. These terms together capture the essential spin and spatial degrees of freedom required to illustrate the magnetosensitivity of radical pairs \cite{adams2018open}.

 In this model, the external magnetic field consists of the static geomagnetic field and the time-dependent component as follows:
 
 \begin{align}
 B ={}& B_0(\sin\theta\cos\phi,\sin\theta\sin\phi,\cos\theta) \nonumber\\
 &+ B_{noise}\cos(\omega t)(\sin\alpha\cos\beta,\sin\alpha\sin\beta,\cos\alpha)
 \label{B}
 \end{align}

where $B_0$ is the magnitude of the Earth’s field, while $B_{\text{noise}}$ and $\omega$ denote the amplitude and angular frequency of the oscillating field, respectively, with $\omega = 2\pi f$, where $f$ is the frequency of the fluctuating field.

The following equation is applied to calculate the electric dipole moment.
\begin{equation}
\mathbf{p} = \sum_i \mathbf{e }\, \mathbf{r_i}
\end{equation}
Where $\mathbf{e }$  is the elementary charge, and $\mathbf{r}$ is the position operator.
In this model, where the Hilbert space is considered spin-based and finite, the position operators are written using ladder operators and the rotating wave approximation. As a result, the spin-orbit evolution is reflected in the magnitude of the dipole moment.
Here,to simplify the calculations, only one nucleus with spin- $\tfrac{1}{2}$ is assumed, and its initial state is fully mixed. Also, the initial state of the two electrons is in the singlet mode, and the orbital angular momentum is modeled as a spin-1 system with three sublevels to allow for transitions between the ground and excited states.
To get a more realistic view of the radical pair system, we applied the effects of the environment, especially in high physiological temperatures that cause quantum coherence decay. The Lindblad master equation \cite {breuer2002theory} was used to implement the dynamics of system-environment interaction.
\begin{equation}
\frac{d}{dt} \rho_s = -\frac{i}{\hbar} [H_s, \rho_s] + \mathcal{L}(\rho_s)
\end{equation}
Where ${L}$ is the Lindblad superoperator, which is expressed as follows:
\begin{equation}
\mathcal{L}(\rho_s) = \Gamma \sum_i \left( A^\dagger_i \rho_s A_i - \frac{1}{2} A^\dagger_i A_i\rho_s - \frac{1}{2} \rho_s A^\dagger_i A_i \right)
 \label{lin}
\end{equation}
Where $\Gamma $ is dissipation rate, and $A $ is the collapse operator, defined as follows:
\begin{equation}
 A_1 = \sigma_1^- \otimes I_2   ~~ \text{and} ~~   A_2 = I_1 \otimes \sigma_2^-        
\end{equation}
where $ I$ is identity matrix and ($\hat{\sigma}^-$) is the energy lowering operators.

\section{Results and discussion}

\begin{figure}
	\centering
	\begin{subfigure}{0.45\columnwidth}
		\centering
		\includegraphics[width=\linewidth]{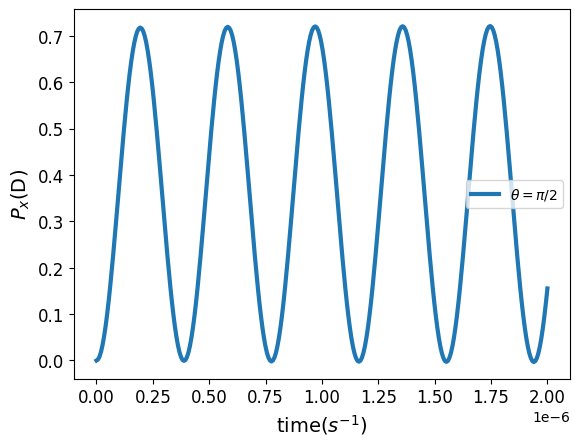}
		\caption{}
		\label{fig:sub1}
	\end{subfigure}
	\hfill
	\begin{subfigure}{0.45\columnwidth}
		\centering
		\includegraphics[width=\linewidth]{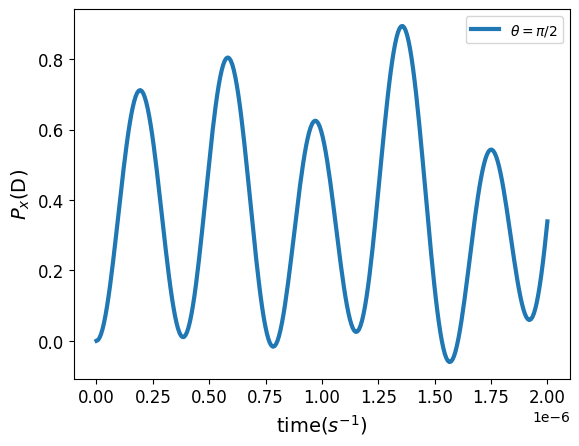}
		\caption{}
		\label{fig:sub2}
	\end{subfigure}
	\caption{Time evolution of the expectation value of electric dipole moment of the radical pair without (a) and with (b) an oscillating magnetic field.}
	\label{withnowith_1}
\end{figure}

With regard to the radical pair mechanism, the interplay between external magnetic fields and  the radical pair spin system plays an important role in determining the dynamics of the system. In our earlier work, the effect of the geomagnetic field on the dynamics of the radical pair dipole moment was investigated and it was shown that changes in the inclination angle and field magnitude can modulate the quantum behavior of the system, results that are in line with experimental observations of orientation-dependent magnetosensitivity in birds. The motivation of the present study was to use quantum modeling to reveal the combined effects of the static Earth’s magnetic field and a radiofrequency (RF) field on the radical pair dipole moment.  
The static magnetic field of 4.6 microtesla is taken as the reference condition on which the RF field is superimposed. Oscillating components can interact with the spin dynamics and possibly lead to phenomena such as resonance effects \cite {Hore2019, Luo2024} that are expected to influence the behavior of the radical pair dipole moment. In the following, the simultaneous effect of the static and time-dependent magnetic field on the expectation value of the electric dipole moment ($P_x$) is investigated under different conditions including various frequencies, intensities, and spatial arrangements of the oscillating field.
To investigate such effects, first the case of a purely static field was compared with the case that an oscillating field was also applied. The results of this comparison are shown in Fig\ref{withnowith_1}; where Fig\ref{withnowith_1} (a) shows evolution of the expectation value of the electrical dipole moment in the absence of the oscillating field (zero magnetic noise) and Fig\ref{withnowith_1} (b) shows the same parameter in the presence of an oscillating field with a frequency of $1.4 MH$ and an intensity of $1\mu T$ and perpendicular to the static field. As can be seen, in the magnetic noise free case, the amplitude of oscillations is constant and a regular periodic behavior is observed. In contrast, with the addition of the oscillating field, the amplitude of the oscillation peaks undergoes significant variations and increases and decreases periodically. This changes indicate that the presence of the oscillating field has caused modulation of the amplitude of oscillations. It should be noted that no noticeable change was observed when the two fields were applied in parallel; therefore, its diagram is not presented.

\begin{figure}
    \centering
    \includegraphics[width=\linewidth]{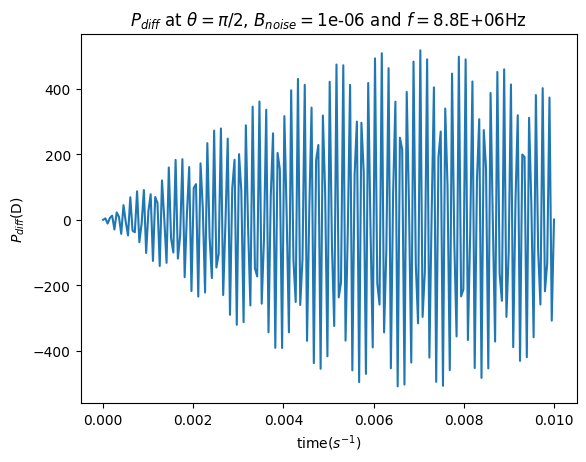}
    \caption{Convergence of the expectation value of $\langle P_{diff}\rangle$ versus the time.}
    \label{converge}
\end{figure}

To reveal the effect of the oscillating magnetic field (as noise) on the time evolution of the system, we define an auxiliary quantity ($P_{diff}$) that represents the comparison of the dipole moment in the two cases of pure static field and the combination of static and oscillating fields. The parameter is introduced as follows:
\begin{equation}
P_{diff} = P_x(B_{noise}=0) - P_x
\end{equation}
Convergence of the expectation value of the defined quantity, is shown at Fig \ref{converge}

\begin{figure}
    \centering
    \includegraphics[width=\linewidth]{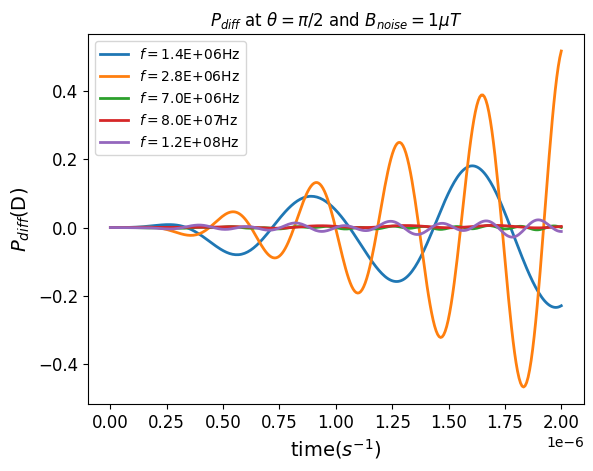}
    \caption{Time evolution of the expectation value ($P_{diff}$) at different frequencies for inclination angle $\theta =\pi/2$ . The oscillating field has a fixed intensity of $1\mu T$ and is applied perpendicular to the static geomagnetic field. At lower frequencies, the system shows clear deviations in the dipole moment, indicating a strong dynamic response of the radical pair. As the frequency increases, this response gradually weakens, consistent with averaging the oscillating field over time. This effect is most pronounced when the oscillating field is perpendicular to the static field. In a case with parallel applied magentic field, changes in frequency, shows almost no deviations.}
    \label{diffvst_freq_90(verand leel)_2}
\end{figure}

Furthermore, the parameter $P_{diff}$ at an inclination angle of  $\theta =\pi/2$ was analyzed in a range of frequencies for the external magnetic field. Here, the intensity of the oscillating field was set to $1\mu T$ and two states perpendicular to the static field and parallel to it are considered. As can be seen in  Fig \ref{diffvst_freq_90(verand leel)_2} , the results show that at low frequencies, the dipole can follow the field changes and as a result, large deviations occur; but at high frequencies, the effect of the field is averaged and the system response will be very small. This is consistent with studies that have reported that oscillating fields within the low MHz spectrum can affect bird navigation \cite{Muheim2023}. On the other hand, the effect of the oscillating field on the radical pair system indicates that the lifetime of the pair was long enough for the spins to have had enough time to perceive the field fluctuations. Geometrically, in a case with the perpendicular field \ref{diffvst_freq_90(verand leel)_2} we see large deviations from a case with no external field, while in the parallel state the interaction is very weak and the changes remain insignificant.

\begin{figure}
	\centering
	\begin{subfigure}[b]{0.45\textwidth}
		\includegraphics[width=\linewidth]{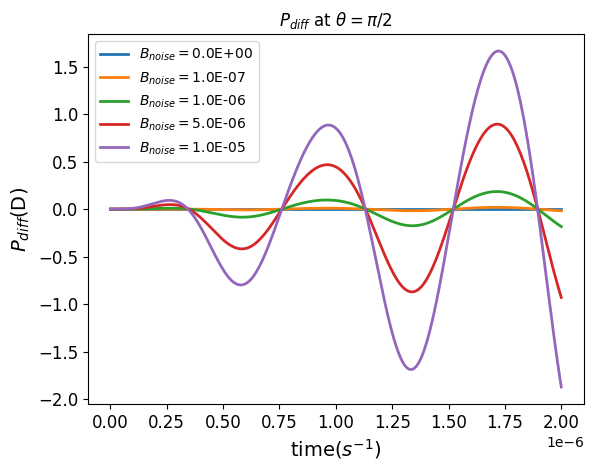}
		\caption{Perpendicular}
	\end{subfigure}
	\hfill
	\begin{subfigure}[b]{0.45\textwidth}
		\includegraphics[width=\linewidth]{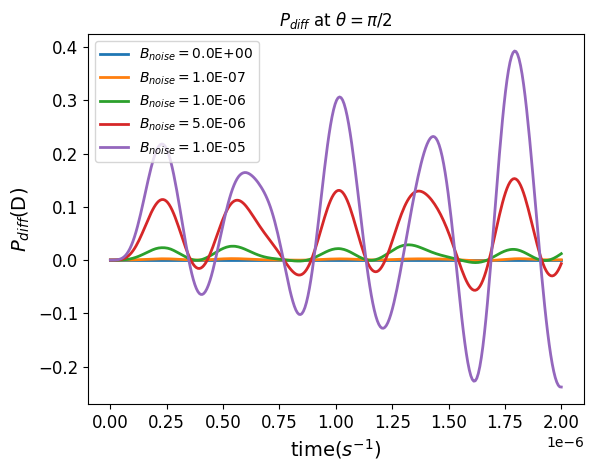}
		\caption{Parallel}
	\end{subfigure}
	\caption{Time evolution of the expectation value of ($P_{diff}$) at intensities for an inclination angle $\theta =\pi/2$ . The oscillating field frequency is fixed at $1.4 MH$ and is applied perpendicular (a) and parallel (b) to the static geomagnetic field. As the field intensity increases, the amplitude of the dipole moment variations grows. This effect is pronounced in the perpendicular configuration.}
	\label{diffvst_intensitty_90(verand leel)_3}
\end{figure}

Now, the effect of the oscillating magnetic field intensity on the time evolution of expectation value of the electric dipole moment, the parameter $P_{diff}$ at an inclination angle of $\theta =\pi/2$ was studied. The frequency of the oscillating field was set to $1.4 MH$ and two modes perpendicular to the static field and parallel to it were considered.  As can be seen in Fig\ref{diffvst_intensitty_90(verand leel)_3}, in general, with increasing intensity of the oscillating field, the amplitude of the oscillations of the electric dipole moment ($P_{diff}$) increases. In the perpendicular state, this effect is very strong and creates large-amplitude oscillations, while in the parallel mode, the interaction is weaker and the oscillations is smaller in the order of one-tenth. Therefore, the intensity and spatial arrangement of the oscillating field are the determining factors in the response amplitude.

\begin{figure}
	\centering
	\begin{subfigure}[b]{0.45\textwidth}
		\includegraphics[width=\linewidth]{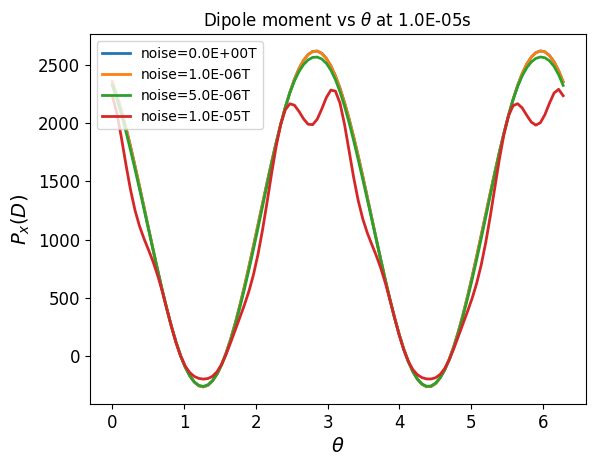}
		\caption{Perpendicular}
	\end{subfigure}
	\hfill
	\begin{subfigure}[b]{0.45\textwidth}
		\includegraphics[width=\linewidth]{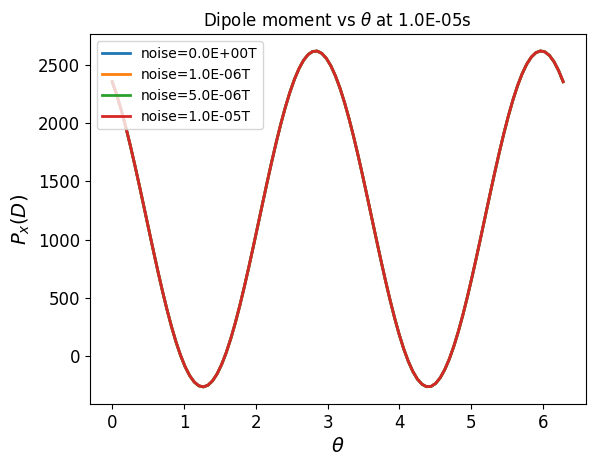}
		\caption{Parallel}
	\end{subfigure}
	\caption{Variations of the electric dipole moment versus inclination angle at different intensities for an oscillating field perpendicular (a) and parallel (b) to the static field. In the vertical configuration, increasing the field intensity leads to significant changes in the amplitude and shape of the dipole moment profile, indicating strong sensitivity to magnetic noise. At low intensities, the response remains close to the static field state.}
	\label{dipplevsteta_intensitty_(verand leel)_5}
\end{figure}

For deeper understanding of the matter, we study the changes of dipole moment versus the inclination angle in the presence of external magnetic field. This is an excellent criterion to show the sensitivity of a radical-paired base avian compass on external noise. In this regard The effect of frequency and intensity of the oscillating field for different inclination angles have been analyzed. Results show that the changes in frequency of the oscillating field almost has no effect on the dependency of dipole moment to the earth magnetic field inclination angle (In cases where the external field is perpendicular and parallel to the Earth static field). On the othe hand, the effect of the magnitude of the external field is shown in Fig \ref{dipplevsteta_intensitty_(verand leel)_5}. Increasing the magnitude of the oscillating field in the perpendicular mode has caused noticeable changes in the amplitude and shape of the curves. At low intensities, the dipole response is almost similar to the magnetic-noise free case, but with increasing the magnitude of the magnetic noise, noticeable deviations in the peaks and a decrease in the regularity of the curves occur. This indicates that the perpendicular oscillating field can effectively change the angular sensitivity pattern of the dipole and introduce noise into the system. In contrast, for the parallel arrangement, the curves are almost perfectly overlapped, and an increase in the oscillating field magnitude has no significant effect on the dipole response. This result indicates that in the parallel mode,magnetic  noise, even at high intensities, is not able to disrupt/change the dependence of the dipole moment to the inclination angle of the Earth magnetic field.

\begin{figure}
    \centering
    \includegraphics[width=\linewidth]{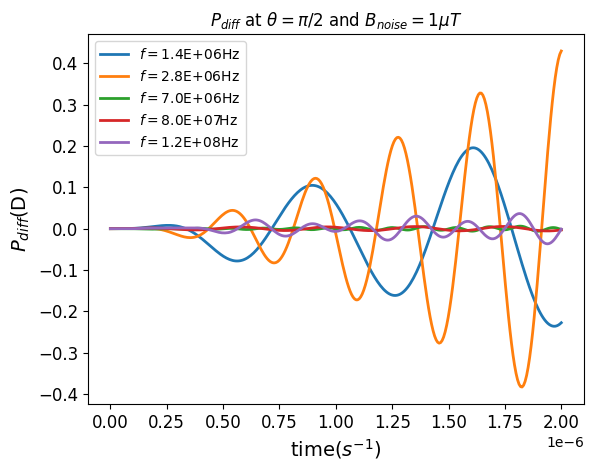}
    \caption{Time evolution of the expectation value of ($P_{diff}$) at different frequencies for an $\theta =\pi/2$ considering the effect of dissipation }
    \label{Noise-frequency-spectrum-dissipation-pi_2_6}
\end{figure}

\begin{figure}
    \centering
    \includegraphics[width=\linewidth]{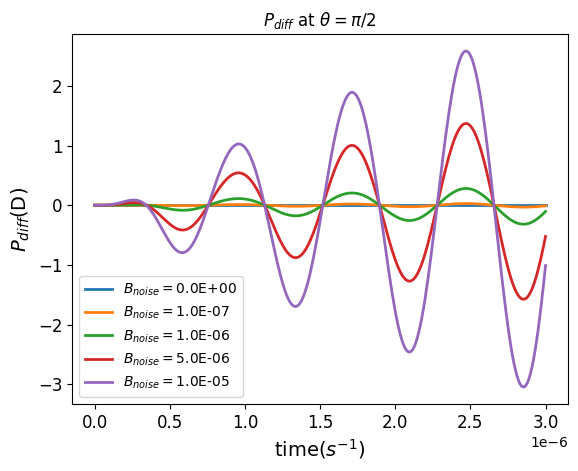}
    \caption{Time evolution of the expectation value ($P_{diff}$) at different intensities for inclination angle $\theta =\pi/2$ considering the effect of dissipation }
    \label{Noise-intensitty-spectrum-dissipation-pi_2_7}
\end{figure}

To consider the effect of environment, study of the time evolution of the radical pair system under the effect of dissipation was carried out by numerically solving the Lindblad master equation. It should be noted that since the results in the parallel mode cannot produce significant changes compared to the case of a purely static magnetic field \cite{xu2013estimating}, in this section, the focus of analyses was placed on the perpendicular mode. Similarly the effect of the frequency (Fig. \ref{Noise-frequency-spectrum-dissipation-pi_2_6}) and intensity (Fig. \ref{Noise-intensitty-spectrum-dissipation-pi_2_7}) of the oscillating magnetic field on the time evolution of expectation value of the electric dipole moment, the auxiliary parameter $P_{diff}$ was analyzed at a inclination angle of $\theta = \pi/2$, respectively. In Fig. \ref{Noise-frequency-spectrum-dissipation-pi_2_6}, the oscillating field magnitude is set to $1\mu T$. As can be seen, in the presence of dissipation and an oscillating field the amplitude of the ($P_{diff}$) oscillations is not strongly dependent on frequency. However, the dissipation effect, amplifies the noise frequency effect, specially in larger frequencies.

In Fig\ref{Noise-intensitty-spectrum-dissipation-pi_2_7}, the oscillating field frequency is set to $1.4 MH$. The results show that in the presence of dissipation, the response ($P_{diff}$) depends on the magnitude of the oscillating field. At low intensities the amplitude of the oscillations is very small and similar to the noiseless case. As intensity increases to about $10^{-5}$ and above, the oscillating field overcomes the dissipation and large-amplitude oscillations appear. Therefore, the magnitude of the magnetic noise is a key factor in determining the amplitude of the response.

\begin{figure}
    \centering
    \includegraphics[width=\linewidth]{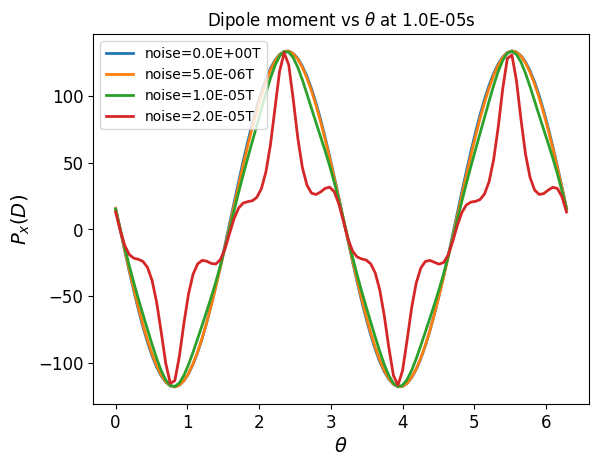}
    \caption{Variations of the electric dipole moment versus inclination angle at different intensities for a perpendicular oscillating field  considering the effect of dissipation. }
    \label{Dipole-vs-theta-intensity-spectrum-dissipation_9}
\end{figure}

Subsequently, at Fig\ref{Dipole-vs-theta-intensity-spectrum-dissipation_9} The effect of the magnetic noise magnitude is shown in a complete spectrum of dipole moment versus inclinationa angle as the criteria of disturbance in the system. As the effect of the magnetic noise frequency is negligible, we do not further discuss its influence on dipole moment changes with inclination angle. Changes the intensity of the oscillating field produces obvious modifications in the angular profile of the dipole moment (Fig\ref{Dipole-vs-theta-intensity-spectrum-dissipation_9}). At low intensities, the response remains almost identical to the static field case, but with increasing intensity, changes in the periodic pattern appear, indicating that a strong oscillating field can significantly alter the dipole response.

\begin{figure}
    \centering
    \includegraphics[width=\linewidth]{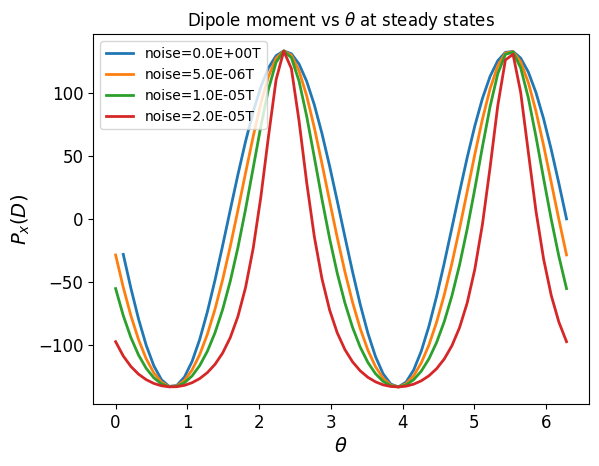}
    \caption{Steady-state electric dipole moment variations versus inclination angle at different intensities for a perpendicular oscillating field  }
    \label{dipplevsteta_intensitty_(noise-intensity-spectrum-steadystate_11}
\end{figure}

Overall, the above findings show that in the presence of dissipation, the frequency of the oscillating field does not have a significant effect on the angular behavior of the dipole moment, while the field intensity plays an important role. Therefore, it can be concluded that the oscillating field intensity is a key parameter in disrupting the dynamics of the radical pair system under real environmental conditions. 
The interaction between system and environment causes the system to go through the decoherence process and inevitably resulting the system to reach to asteay state.  In order to investigate this issue, the changes in the the electric dipole moment with respect to the inclination angle in the steady state were studied. Examining the dependence of the dipole moment on the intensity of the oscillating field in the steady state shows that this parameter plays a prominent role in the angular pattern (Fig\ref{dipplevsteta_intensitty_(noise-intensity-spectrum-steadystate_11}). In such a way that increasing it changes the shape of the profile and reduces  its symmetry. Therefore, the intensity of the field can be considered as a key factor in determining the stability and quantum behavior of the dipole moment.

\subsection{Effect of the oscillating field oriented 24 degrees relative to the Geomagnetic field}

\begin{figure}
	\centering
	\begin{subfigure}[b]{0.45\textwidth}
		\includegraphics[width=\linewidth]{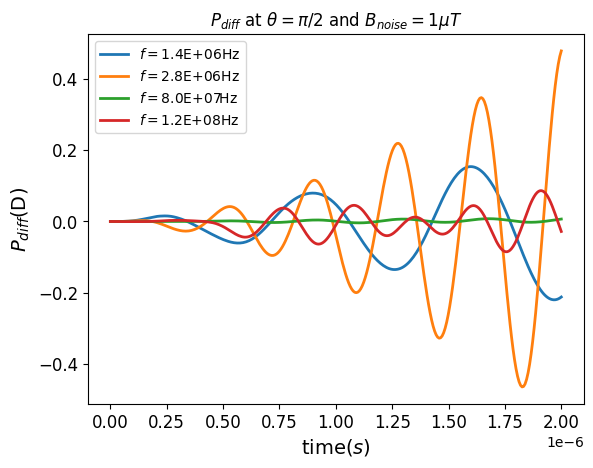}
		\caption{}
	\end{subfigure}
	\hfill
	\begin{subfigure}[b]{0.45\textwidth}
		\includegraphics[width=\linewidth]{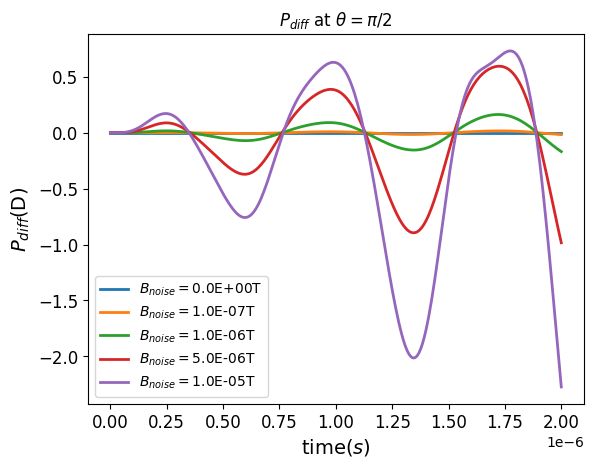}
		\caption{}
	\end{subfigure}
	\caption{Time evolution of the expectation value of ($P_{diff}$) at different frequencies (a) and intensities (b) for the oscillating field oriented $24^\circ{}$ relative to the Geomagnetic field at inclination angle $\theta =\pi/2$}
	\label{fig 13}
\end{figure}

Experimental Report \cite{Ritz2004} discusses that birds navigation shows the most perturbation and disorientation at an angle of 24 degrees between an oscillating and a static magnetic fields. We have likewise investigated the behavior and changes in the electric dipole moment at this specific angle. Comparison of the time evolution of the electric dipole moment in the two configurations of perpendicular and $24^\circ{}$ angles shows that the system exhibits greater sensitivity to frequency variations at $24^\circ{}$. In this angle, the system dependency and sensitivity to noise frequency increases and follows a different pattern than a perpendicular noise (Fig\ref{fig 13}a). Also, High intensities show a complete disturbance at this angle. As Fig\ref{fig 13}b shows, oscillating noise causes remarkable changes in radical-pair dipole moment, and losing the symmetry in dipole changes may also play a substantial role in birds' disorientation.

\begin{figure}
	\centering
	\begin{subfigure}[b]{0.45\textwidth}
		\includegraphics[width=\linewidth]{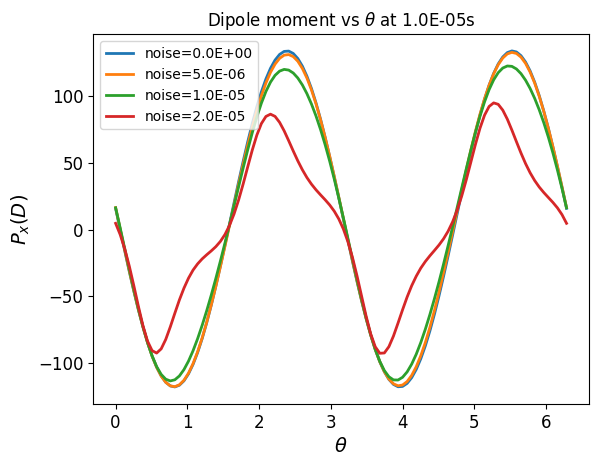}
		\caption{}
	\end{subfigure}
	\hfill
	\begin{subfigure}[b]{0.45\textwidth}
		\includegraphics[width=\linewidth]{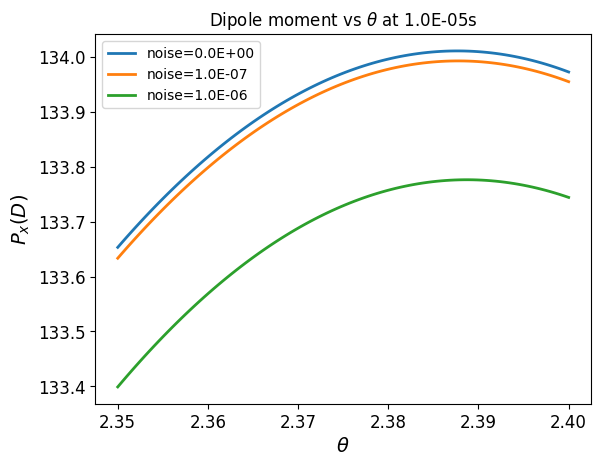}
		\caption{}
	\end{subfigure}
	\caption{Variations of the electric dipole moment versus inclination angle at different intensities (microtesla (a) and nanotesla (b)) for the oscillating field oriented $24^\circ{}$ relative to the Geomagnetic field considering the effect of dissipation at $\theta =\pi/2$. At this angle the system is sensitive even to the very low intensities of magnetic noise.}
	\label{Fig 14}
\end{figure}

Such differences in the electric dipole moment plots versus the inclination angle between the perpendicular and $24^\circ{}$ configurations were also observed in the ideal case (not shown here), but the inclusion of dissipative effects of the environment make these differences more pronounced. As shown in Fig\ref{Fig 14}a, at an angle of $24^\circ{}$, the system produces distinct responses even at the lowest field intensities even at $\sim$ 100 nanotesla (Fig\ref{Fig 14}b), and the curves do not overlap, whereas in the perpendicular case, these responses are almost completely negligible. Therefore, at $24^\circ{}$, and when environmental effects are taken into account, the effect of magnetic noise increases and further perturbs the dynamics of the system, which can lead to significant disorientation.
\begin{figure}
    \centering
    \includegraphics[width=\linewidth]{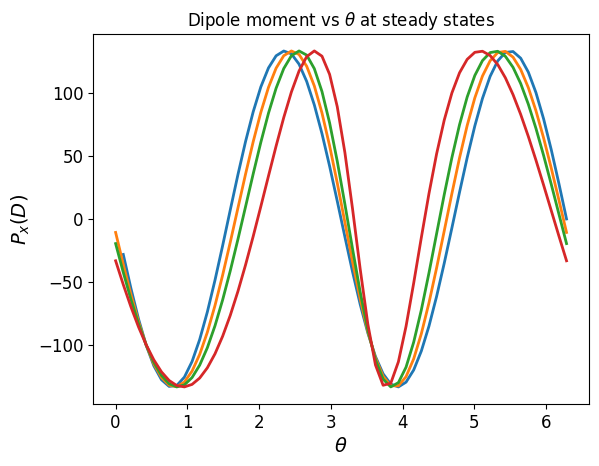}
    \caption{Steady-state electric dipole moment variations versus inclination angle at different intensities for the oscillating field oriented $24^\circ{}$ relative to the Geomagnetic field at $\theta =\pi/2$}
    \label{Fig 15_noise-intensity-spectrum-steadystate}
\end{figure}

 In the steady-state regime at an angle of $24^\circ{}$ (Fig\ref{Fig 15_noise-intensity-spectrum-steadystate}), a clear shift in the peak positions is observed at different magnetic field intensities. Comparison with the perpendicular mode indicates that, at this angle, environmental and magnetic noise introduce stronger perturbations to the avian orientation system.The significant disruption at $24^\circ{}$ in our results is consistent with experimental findings regarding bird disorientation at the same angle.
Overall, comparing the results for the parallel, perpendicular, and intermediate configurations shows that the magnitude or pattern of the disturbance caused by magnetic noise depends on the relative fields orientation \cite{Ritz2004, Muheim2023}.This directional dependence further supports the involvement of the radical-pair mechanism in magnetoreception process.

\section{ Concluding remarks}

In this study, we aimed to gain an in-depth insight of the magnetic sense of organisms under an oscillating magnetic field superimposed on a static geomagnetic field within a quantum framework. The changes in the electric dipole moment of the radical pair due to these combined fields have been investigated by numerical solution. Calculations show that the electric dipole moment depends on the characteristics of the oscillating field, such as frequency, intensity, and its relative angle with the static field, even in the presence of dissipative effects. In general, based on the results, it can be concluded that radiofrequency fields in the studied range culminate in a change in the pattern that is strongly dependable to the intensity of the magnetic-noise and also its geometric angle relative to the Erth's magnetic field. The birds behaviorial studies are in complete agreement with our findings. These results support radical pair-based magnetoreception and pave the way for the development of technologies such as quantum biosensors and magnetic navigation inspired through magnetoreception.

\bibliographystyle{unsrtnat}
\bibliography{inorder.bib}
\end{document}